\documentclass[aps,prd,floatfix,twocolumn,amsmath,amssymb,nofootinbib,preprintnumbers]{revtex4-1}
\usepackage{amsthm,color,latexsym,graphicx,array,multirow,epstopdf,slashed}
\usepackage[colorlinks=true,linkcolor=darkblue,citecolor=darkblue,urlcolor=darkblue]{hyperref}
\definecolor{darkblue}{rgb}{0,0,0.5}
\definecolor{red}{rgb}{0.75,0,0}

\newcommand{\gev}{\mathrm{GeV}}
\newcommand{\tev}{\mathrm{TeV}}
\newcommand{\met}{\slashed{E}_T}

\begin{document}
\preprint{EFI-15-3}
\title{Neutralinos in Vector Boson Fusion at High Energy Colliders}

\author{Asher Berlin\vspace{2ex}}
\email{berlin@uchicago.edu}
\affiliation{Department of Physics, Enrico Fermi Institute, and Kavli Institute for Cosmological Physics, University of Chicago, Chicago IL, 60637 \vspace{2ex}}

\author{Tongyan Lin}
\email{tongyan@kicp.uchicago.edu}
\affiliation{Department of Physics, Enrico Fermi Institute, and Kavli Institute for Cosmological Physics, University of Chicago, Chicago IL, 60637 \vspace{2ex}}

\author{Matthew Low}
\email{mattlow@uchicago.edu}
\affiliation{Department of Physics, Enrico Fermi Institute, and Kavli Institute for Cosmological Physics, University of Chicago, Chicago IL, 60637 \vspace{2ex}}

\author{Lian-Tao Wang}
\email{liantaow@uchicago.edu}
\affiliation{Department of Physics, Enrico Fermi Institute, and Kavli Institute for Cosmological Physics, University of Chicago, Chicago IL, 60637 \vspace{2ex}}

\date{\today}

\begin{abstract} 
Discovering dark matter at high energy colliders continues to be a compelling and well-motivated possibility.  Weakly interacting massive particles are a particularly interesting class in which the dark matter particles interact with the standard model weak gauge bosons.  Neutralinos are a prototypical example that arise in supersymmetric models.  In the limit where all other superpartners are decoupled, it is known that for relic density motivated masses, the rates for neutralinos
are too small to be discovered at the Large Hadron Collider (LHC), but that they may be large enough to observe at a 100 TeV. In this work we perform a careful study in the vector boson fusion channel for pure winos and pure higgsinos.  We find that given a systematic uncertainty of 1\% (5\%), with 3000 fb$^{-1}$, the LHC is sensitive to winos of 240 GeV (125 GeV) and higgsinos of 125 GeV (55 GeV).  A future 100 TeV collider would be sensitive to winos of 1.1 TeV (750 GeV) and higgsinos of 530 GeV (180 GeV) with a 1\% (5\%) uncertainty, also with 3000 fb$^{-1}$.
\end{abstract}

\maketitle

%%%%%%%%%%%%%%%%%%%%%%%%%%%%%%%%%%%%%%%%%%%%%%%%%%%%%%%%%%%%%%%%%%%%%%%%%%%%%%%%%%%%%%%%%%%%%%%%%%%%%%%%%%%%%%%%%%%%%%%%
\section{Introduction}\label{sec:introduction}

Among the possibilities of new physics at the Large Hadron Collider (LHC), observing dark matter is certainly one of the most exciting.  Despite the fact that dark matter makes up a sizable fraction of the energy budget of the universe, we remain in the dark regarding its identity.  While there are many candidates, weakly interacting massive particles (WIMPs) are particularly tantalizing because of their weak-scale annihilation cross-section and the potential to see signals in collider experiments, direct detection, and indirect detection.  Because direct and indirect detection are subject to large astrophysical uncertainties, producing dark matter at colliders seems to be especially crucial in discerning its properties.

Many approaches to sweep through dark matter parameter space have been proposed, including the use of effective operators~\cite{Goodman:2010ku,Zhou:2013raa,Cao:2009uw,Goodman:2010yf,Bai:2010hh,Kopp:2011eu,Rajaraman:2011wf} and simpified models~\cite{An:2012va,An:2012ue,An:2013xka,Malik:2014ggr,deSimone:2014pda}. Generically these approaches characterize the process $pp \to \tilde{\chi}\tilde{\chi}$ where $\tilde{\chi}$ is the dark matter particle and is observed as missing energy.  An observation then requires a detectable particle to be radiated off of the initial state.\footnote{One can also consider radiation off of the dark matter itself which can lead to qualitatively different signals~\cite{Lin:future}.}  These ``mono-X'' signatures include monojet~\cite{Beltran:2010ww,Fox:2011pm}, mono-photon~\cite{Gershtein:2008bf,Fox:2011fx}, mono-$Z$~\cite{Petriello:2008pu,Carpenter:2012rg}, mono-$W$~\cite{Bai:2012xg}, mono-Higgs~\cite{Petrov:2013nia,Carpenter:2013xra,Berlin:2014cfa}, mono-$b$~\cite{Lin:2013sca,Artoni:2013zba,Abdallah:2014hon}, and mono-top~\cite{Lin:2013sca,Artoni:2013zba,Abdallah:2014hon}.

\begin{figure}[b]
  \includegraphics[width=0.49\textwidth]{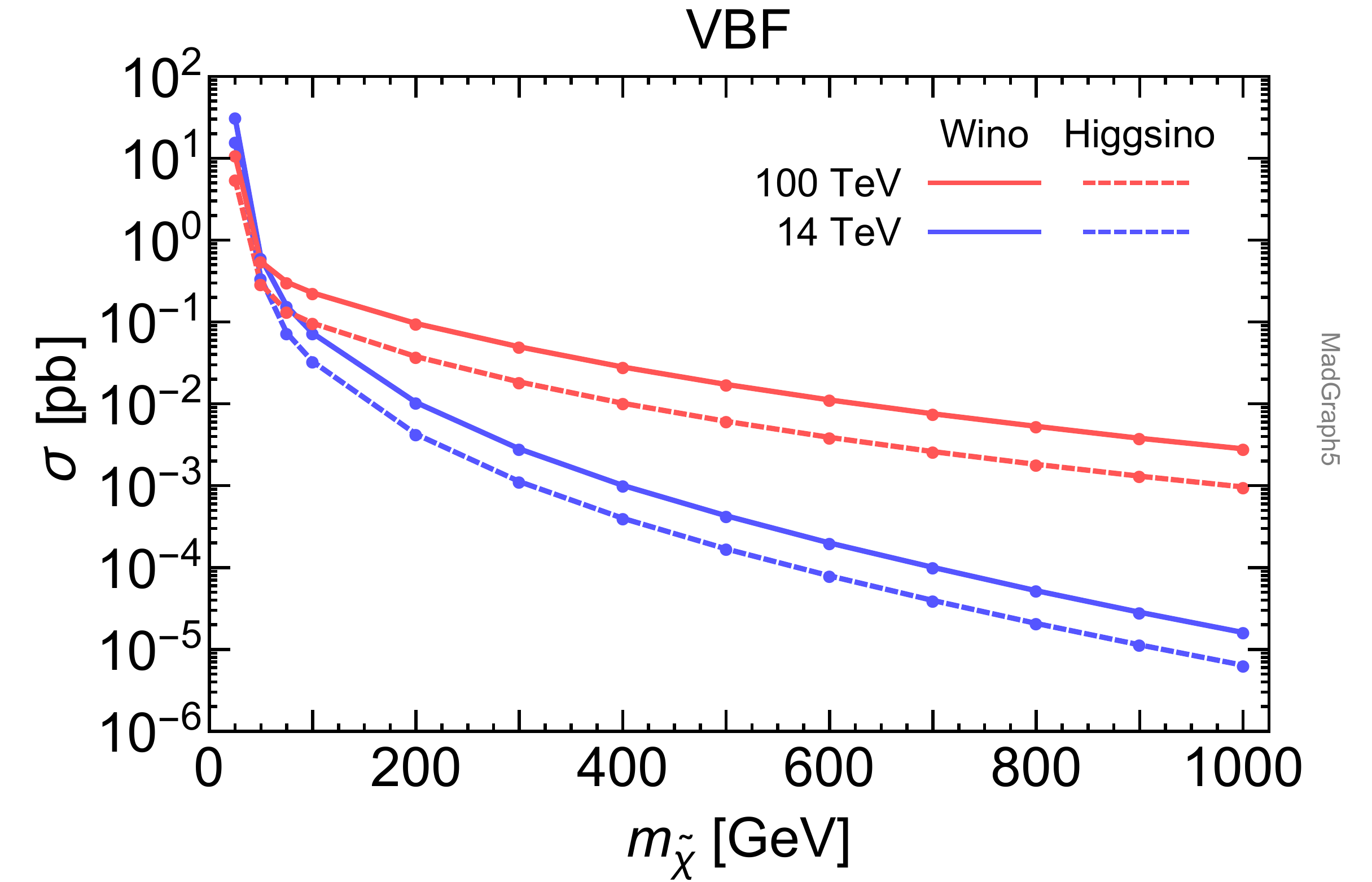}
  \caption{Production cross sections where for 14 (100) TeV jets have $p_T>50$ GeV, are separated by $\Delta\eta>$ 3.5 (4.0), and a dijet invariant mass $>$ 100 GeV (500 GeV).  The missing energy requirement is $>$ 100 GeV (500 GeV).}
  \label{fig:production}
\end{figure}

In a UV-complete model of WIMP dark matter the simplest parameterization is to add an electroweak multiplet to the standard model.  The minimal supersymmetric standard model (MSSM) already provides the canonical examples of new fermion multiplets in the form of neutralino dark matter: a singlet (bino), doublet (higgsino), and triplet (wino). In this work we study the higgsino and wino, however, the results are completely general  and not intrinsically supersymmetric.\footnote{To leading order the pure bino has no electroweak interactions and overcloses the universe when produced thermally, so we do not consider it further.}

In order to get the correct relic density, the mass for winos needs to be $\sim 3~\tev$ and for higgsinos needs to be $\sim 1~\tev$~\cite{Hisano:2006nn,Cirelli:2007xd}.  While it is well known that non-thermal production can allow for different masses to saturate the relic density, the thermal value is a useful benchmark for which to aim.  From the supersymmetric point of view, we also take the most conservative approach by assuming that the dark matter multiplet is only produced directly rather than at the end of decay chains of heavier sparticles.  In this scenario, it has been shown that the LHC will not be able to reach the thermal relic region~\cite{Low:2014cba} (see also~\cite{Cirelli:2014dsa}).  Generally the most sensitive search is the monojet final state.  Due to electroweak symmetry breaking, quantum corrections split the masses of the charginos and neutralinos such that the decays of charginos can result in the disappearing tracks signature which, for small enough splitting, can be much more sensitive than monojet~\cite{Low:2014cba}.  It is still important, however, to understand and evaluate different search channels as their conclusions are complementary and can be affected by very different systematics.  

In this vein, we study the reach for winos and higgsinos in the vector boson fusion (VBF) channel at the 14 TeV LHC and at a proposed future 100 TeV $pp$ collider.\footnote{See~\cite{Hook:2014rka,Cirelli:2014dsa,Bramante:2014tba,Curtin:2014jma,Craig:2014lda,Ismail:future} for other 100 TeV studies relating to vector boson fusion.} The importance of the vector boson fusion signature has long been recognized in the study of strong electroweak symmetry breaking~\cite{Bagger:1993zf,Bagger:1995mk} and more recently, in dark matter searches~\cite{Datta:2000ja,Datta:2001cy,Datta:2001hv,Choudhury:2003hq,Konar:2003pn,Giudice:2010wb,Dutta:2012xe,Delannoy:2013ata,Dutta:2013gga,Delannoy:2013dla}.   In particular, electroweakinos can be produced via VBF, with a signature of forward jets and missing energy. 
We determine the optimized sensitivity of this search  to wino and higgsino dark matter, and also present results that can be applied to any model with new physics in VBF and missing energy.

Note that similar work has been done in~\cite{Cirelli:2014dsa,Delannoy:2013ata}; this work is complementary (and agrees broadly with the conclusions of~\cite{Cirelli:2014dsa}).  In our work we perform matching for additional jets.  This is especially important for the VBF signal because it suffers from large uncertainties from factorization scale choice~\cite{Han:2009em}.  Jet matching can stabilize uncertainties in the parton shower.  Additionally we study the impact of future detector design on this search.

The details of the paper are as follows.  In Section~\ref{sec:analysis} we describe the analysis used to isolate the VBF signal.  Section~\ref{sec:simulation} presents the details of the simulation implemented to compute the signal and backgrounds.  The results for the pure wino and pure higgsino are shown in Section~\ref{sec:results} and for a generalized dark matter model in Section~\ref{sec:general}.  Finally, Section~\ref{sec:conclusions} contains our conclusions.

%%%%%%%%%%%%%%%%%%%%%%%%%%%%%%%%%%%%%%%%%%%%%%%%%%%%%%%%%%%%%%%%%%%%%%%%%%%%%%%%%%%%%%%%%%%%%%%%%%%%%%%%%%%%%%%%%%%%%%%%
\section{Analysis and Simulation}

\begin{table} [b]
  \begin{tabular}{| c | cc | cc |}
    \hline
    \multirow{2}{*}{Cut}          & \multicolumn{2}{c|}{14 TeV}     & \multicolumn{2}{c|}{100 TeV}   \\
                                  & ~~Wino~~        & ~Higgsino~    & ~~Wino~~      & ~Higgsino~     \\ \hline \hline
    $n_{\text{jet}}$              & 2      & 2      & 2      & 2    \\
    $|\eta(j)|$                   & 5      & 5      & 7      & 7    \\
    $p_T(j_{\text{tag}})$ (GeV)   & 45     & 45     & 75     & 50   \\
    $\Delta\eta(j_1,j_2)$         & 3.75   & 3.75   & 4.25   & 4.25 \\
    $\Delta\phi(j_1,j_2)$         & 2      & 2      & 2      & 3    \\
    $M(j_1,j_2)$ (TeV)            & 2      & 1      & 10     & 5    \\
    $\met$ (GeV)                  & \multicolumn{2}{c|}{400 -- 700} & \multicolumn{2}{c|}{1100 -- 2500} \\
    $p_T(j_{\text{veto}})$ (GeV)  & 45     & 45     & 50     & 50   \\
    $p_T(e,\mu)$ (GeV)            & 20     & 20     & 20     & 20   \\
    $p_T(\tau)$ (GeV)             & 30     & 30     & 40     & 40   \\
    $\eta(e)$                     & 2.5    & 2.5    & 2.5    & 2.5  \\
    $\eta(\mu)$                   & 2.1    & 2.1    & 2.1    & 2.1  \\
    $\eta(\tau)$                  & 2.3    & 2.3    & 2.3    & 2.3  \\
    \hline
  \end{tabular}
  \caption{Cuts used for each signal for $3000$ fb$^{-1}$ of integrated luminosity.  The $\met$ cut is optimized for each mass point.  Details are described in Section~\ref{sec:analysis}.}
  \label{table:cuts}
\end{table}
%

% REPLACE THESE
%
\begin{figure*}[tb]
  \includegraphics[width=0.49\textwidth]{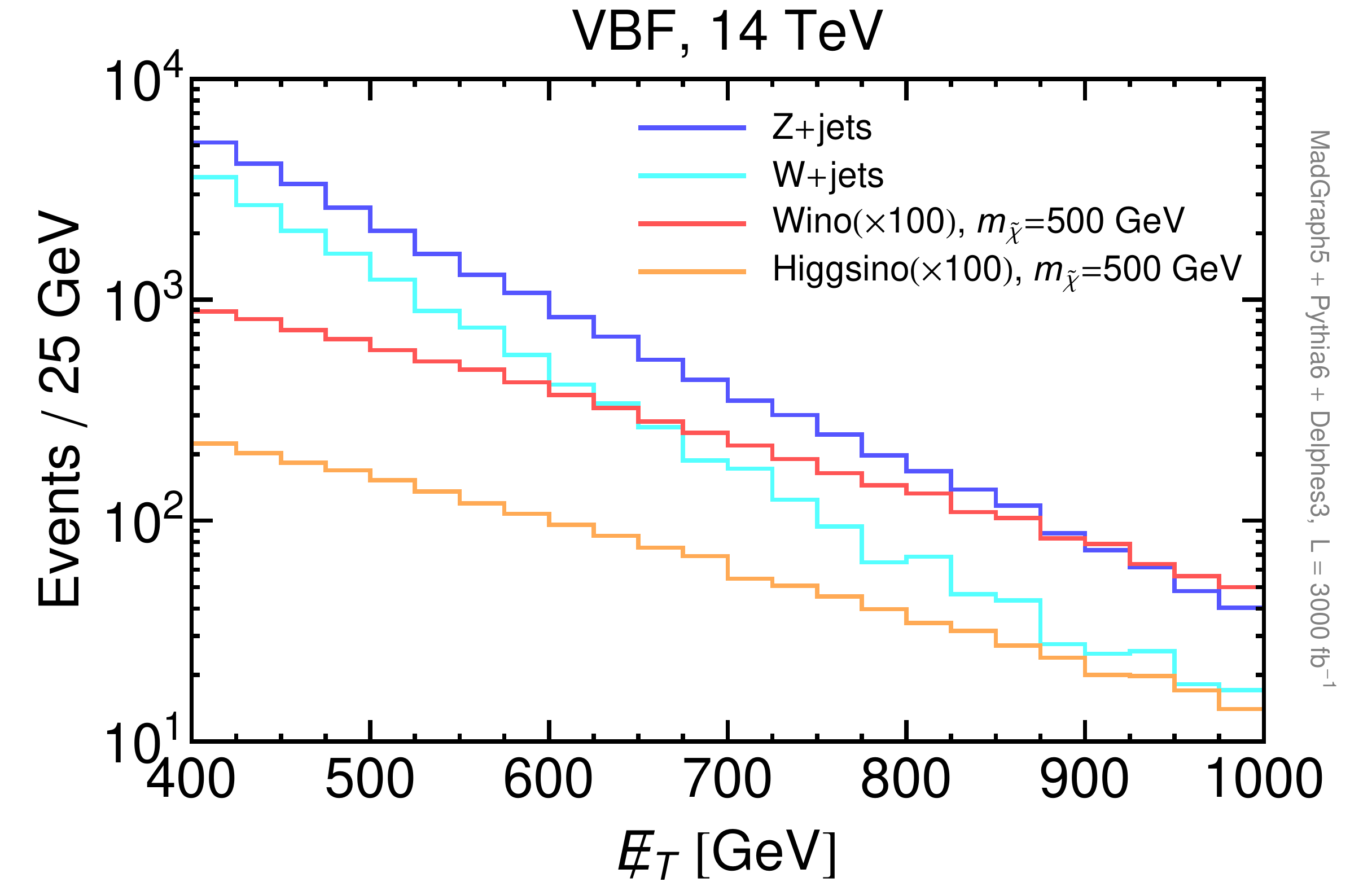} 
  \includegraphics[width=0.49\textwidth]{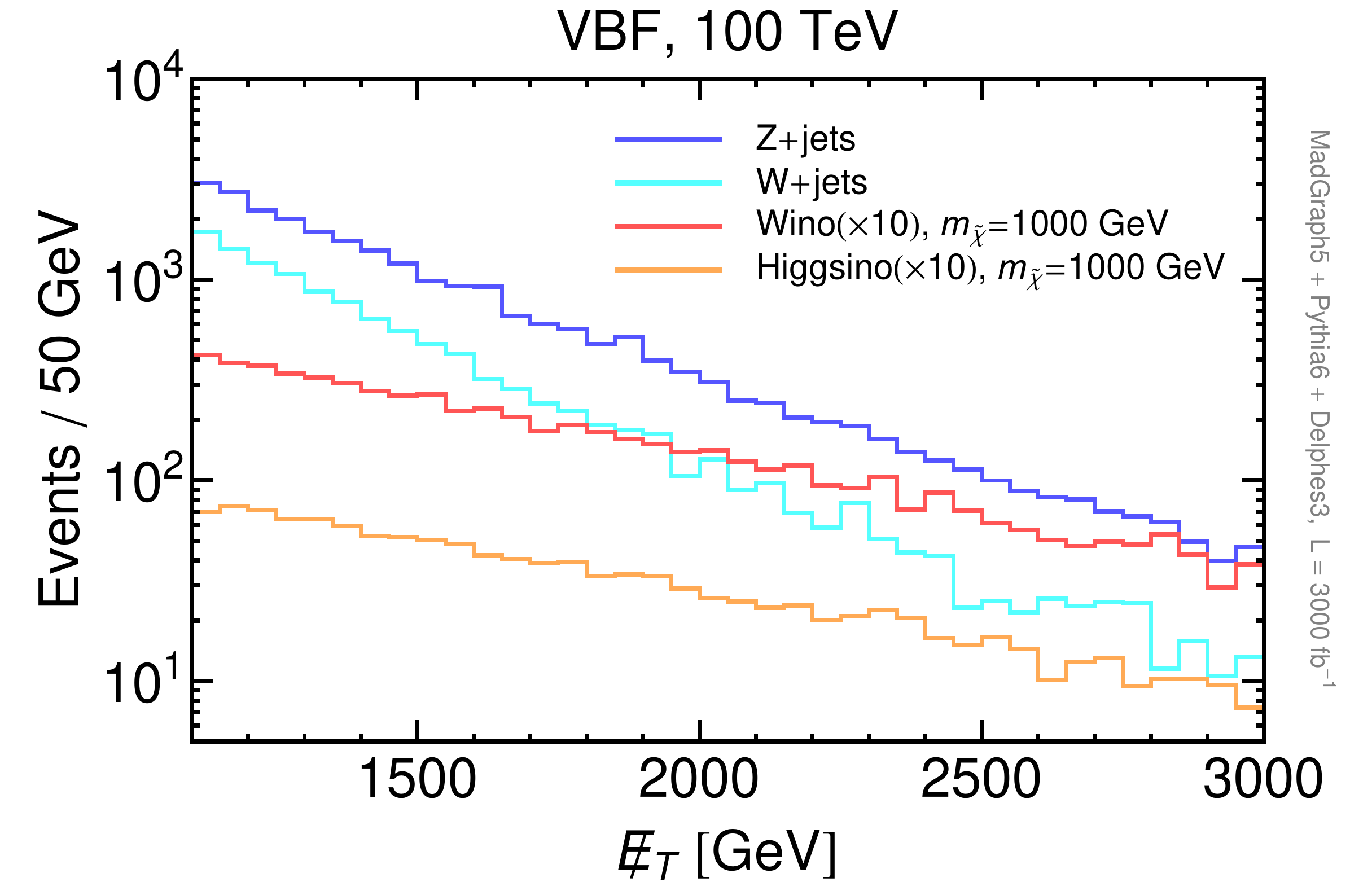}
  \caption{Distributions of missing transverse energy for 14 TeV (left) and 100 TeV (right).  All the cuts (averaged between the wino and higgsino values) of Table~\ref{table:cuts} are applied except for the missing transverse energy cut.  The signal rates are multiplied by a factor of 100 and 10 at 14 and 100 TeV, respectively.}
  \label{fig:met}
\end{figure*}
%

%%%%%%%%%%%%%%%%%%%%%%%%%%%%%%%%%%%%%%%%%%%%%%%%%%%%%%%%%%%%%%%%%%%%%%%%%%%%%%%%%%%%%%%%%%%%%%%%%%%%%%%%%%%%%%%%%%%%%%%%
\subsection{Analysis Details} \label{sec:analysis}

Although there are many possible variations on the VBF signal, the general strategy is to tag two forward jets and then look for the new particles in the central region.  The presence of these new particles may be indicated by missing energy, or by decay products like leptons, which can be helpful when considering compressed spectra. Again, we take the conservative point of view and assume that both charginos and neutralinos result solely in missing energy rather than considering the case where one is able to tag on some of the chargino decay products.

The wino and higgsino cross-sections in vector boson fusion are plotted in Figure~\ref{fig:production}. In practice, including additional cuts improves the significance and will be described below.

The largest backgrounds are $Z(\nu\nu)$+jets, $W(\ell\nu)$+jets, $t\bar{t}$, and QCD multijet.  The $W(\ell\nu)$+jets background can contribute when the $W$ decays leptonically and the lepton fails to be tagged because it is outside the detector acceptance, not isolated, or too soft.  Similarly the $t\bar{t}$ background is primarily from the fully leptonic decay where both leptons are missed.  The multijet background arises primarily from the mismeasurement of a jet which mimics the jets and missing energy signal.  Note that the vector boson + jets backgrounds can be produced at $\mathcal{O}(\alpha_w \alpha_s^2)$ and $\mathcal{O}(\alpha_w^3)$ and both contribute comparably in our phase space.

We employ the following analysis.  The values for the cuts used in each signal are listed in Table~\ref{table:cuts}.
\begin{itemize}

  \item There are required to be $\geq n_{\text{jet}}$ jets, where jets are defined to have $p_T(j) > 45~\gev$ and pseudorapidity $< |\eta(j)|$.
  
  \item The two hardest jets, $j_1$ and $j_2$, are required to have $p_T(j_{1,2}) > p_T(j_{\text{tag}})$, and $j_1$ and $j_2$ are required to be opposite, {\it i.e.} $\eta(j_1) \cdot \eta(j_2) < 0$, to be separated in $\eta$ by more than $ \Delta\eta(j_1,j_2)$, separated in $\phi$ by less than $ \Delta\phi(j_1,j_2)$, and have an invariant mass larger than $ M(j_1,j_2)$.
  
  \item There must be missing transverse energy ($\met$) in the event.  This cut depends strongly on the dark matter mass and is optimized per mass point.

  \item Central jets are vetoed, where a central jet, $j$, is defined to have $p_T(j) > p_T(j_{\text{veto}})$ and be between the two forward jets $\eta_{\text{min}} < \eta(j) < \eta_{\text{max}}$, where $\eta_{\text{min}} \equiv \min(\eta(j_1),\eta(j_2))$ and $\eta_{\text{max}} \equiv \max(\eta(j_1),\eta(j_2))$.

  \item Identified leptons are vetoed, where Table~\ref{table:cuts} describes the minimum $p_T$ and maximum $|\eta|$ requirements on the leptons.  Standard isolation criteria are applied and hadronic taus are tagged with an efficiency of 50\%.
    
\end{itemize}

Missing energy distributions are shown in Figure~\ref{fig:met}. The $\met$ spectrum for wino and higgsino falls more slowly compared to the dominant backgrounds due to the production of heavy on-shell particles. Since the missing energy spectrum of neutralinos is directly tied to their masses,  for each mass point the $\met$ cut is separately optimized. We optimized over all the cuts and find that, with the exception of $\met$, the cuts are not strongly sensitive to the neutralino mass.

QCD multijet is a potential background due to mismeasuring a jet resulting in seemingly large missing energy.  As these events tend to come from dijet events, the azimuthal cut between the tagging jets largely removes this configuration which is why we neglect this background.  In the signal the jets recoil from the neutralinos and have no preference to be back-to-back.\footnote{The azimuthal cut is not entirely uncorrelated with the other cuts.  Decreasing $\Delta\phi(j_1,j_2)$ tends to make the leading jets align more which preferentially selects events with higher $\met$.}

The significance for each mass point is computed as
\begin{equation}
  \text{Significance} =  
  \frac{S}{\sqrt{B + \gamma_B^2 B^2 + \gamma_S^2 S^2}} ,
  \label{eq:significance}
\end{equation}
where $S$ is the number of signal events, $B$ is the number of background events, and $\gamma_{B,S}$ is the systematic uncertainty on the background and signal, respectively.  The analysis in~\cite{Chatrchyan:2014tja} employs a similar event selection and has a systematic uncertainty of 15\%.  If one over-optimistically scales this with luminosity, the systematics at $\mathcal{L} = 3000~\text{fb}^{-1}$ would be $\approx 1.5\%$.  In this work $\gamma_B$ is varied between $1 - 5\%$ and $\gamma_S = 10\%$ is fixed.  While even this may be too hopeful, it provides a useful benchmark.

%%%%%%%%%%%%%%%%%%%%%%%%%%%%%%%%%%%%%%%%%%%%%%%%%%%%%%%%%%%%%%%%%%%%%%%%%%%%%%%%%%%%%%%%%%%%%%%%%%%%%%%%%%%%%%%%%%%%%%%%
\subsection{Simulation Details} \label{sec:simulation}

The signal spectrum is generated using SUSPECT2~\cite{Djouadi:2002ze} with $\tan\beta=20$.  Since we decouple all superpartners (except for the electroweakino multiplet under consideration), varying $\tan \beta$ has little effect on our study. Signal and background events are generated using MadGraph 5 v2.1.1~\cite{Sjostrand:2006za} and showered and hadronized using Pythia 6 v2.3.0~\cite{Alwall:2011uj}.  We use Delphes 3 v3.1.2~\cite{deFavereau:2013fsa} as the detector simulation, using modified versions of the Snowmass cards~\cite{Cohen:2013xda,Avetisyan:2013dta,Anderson:2013kxz} and Fastjet v3.0.6~\cite{Cacciari:2011ma} to cluster anti-$k_T$ jets with a radius of $R=0.5$~\cite{Cacciari:2008gp}.  

For the signal we generate both the $\mathcal{O}(\alpha_w^3)$ and the $\mathcal{O}(\alpha_w \alpha_s^2)$ contributions.  We do the same for the $Z(\nu\nu)$+jets and $W(\ell\nu)$+jets backgrounds.  We also simulate the leptonic $t\bar{t}$ background.  We neglect the hadronic $t\bar{t}$ backgrounds because the missing energy cut effectively removes them.  Similarly we do not simulate the multijet background because the missing energy and $\Delta\phi$ cuts make it negligible.  All events were matched up to one additional jet using the MLM shower-$k_T$ matching scheme. The events used were weighted between samples with different generator-level missing energy cuts.

We do not apply $k$-factors to our signal or background events. The known $k$-factors for $2 \to 2$ processes will differ from the $k$-factors for the $2 \to 4$ processes considered in this study, and the signal cross section diminishes quickly as a function of increasing dark matter mass. Hence, the omission of $k$-factors is estimated to be a minor effect.

In anticipation of future detector design we extended the coverage for jets up to $|\eta| = 7$ (from $|\eta|=5$).  This modification is useful to evaluate the impact of detector design on reach.  It turns out, as will be discussed in Section~\ref{sec:results}, changing the detector extent in $\eta$ does not affect the reach in this channel.

In our studies we neglect the impact of pileup.  There have been early studies indicating that the inclusion of pileup in simulations changes the efficacy of the central jet veto~\cite{Gianotti:2002xx}, however, new developments in pileup removal are expected to significantly mitigate these effects~\cite{Krohn:2013lba,Berta:2014eza,Cacciari:2014gra,Bertolini:2014bba}.

\begin{table}
  \begin{tabular}{| c | c c | c c |}
  \hline
  \multirow{2}{*}{\textbf{14 TeV}} & \multicolumn{2}{c|}{~~~$95\%$~~~}  & \multicolumn{2}{c|}{~~~~$5\sigma$~~~} \\ 
                                   & ~~1\%~~  & ~~5\%~~   & ~~1\%~~  & ~~5\%~~   \\ \hline\hline
  ~~Wino~~(GeV)                         & 240  & 125   & 135  & 50  \\
  ~~Higgsino~~(GeV)                     & 125  & 55    & 70   & 40  \\ 
  \hline
  \end{tabular}
  \caption{Mass reach for winos and higgsinos at 14 TeV with 3000 fb$^{-1}$ of integrated luminosity.}
  \label{table:14TeVresults}
\end{table}
\begin{table}
  \begin{tabular}{| c | c c | c c |}
  \hline
  \multirow{2}{*}{\textbf{100 TeV}} & \multicolumn{2}{c|}{~~~$95\%$~~~}  & \multicolumn{2}{c|}{~~~~$5\sigma$~~~} \\ 
                                   & ~~1\%~~  & ~~5\%~~   & ~~1\%~~  & ~~5\%~~   \\ \hline\hline
  ~~Wino~~(GeV)                         & 1100 & 750 & 600 & 220  \\
  ~~Higgsino~~(GeV)                     & 530 & 180 & 180 & 50  \\ 
  \hline
  \end{tabular}
  \caption{Mass reach for winos and higgsinos at 100 TeV with 3000 fb$^{-1}$ of integrated luminosity.}
  \label{table:100TeVresults}
\end{table}
\begin{figure*}[t]
  \includegraphics[width=0.49\textwidth]{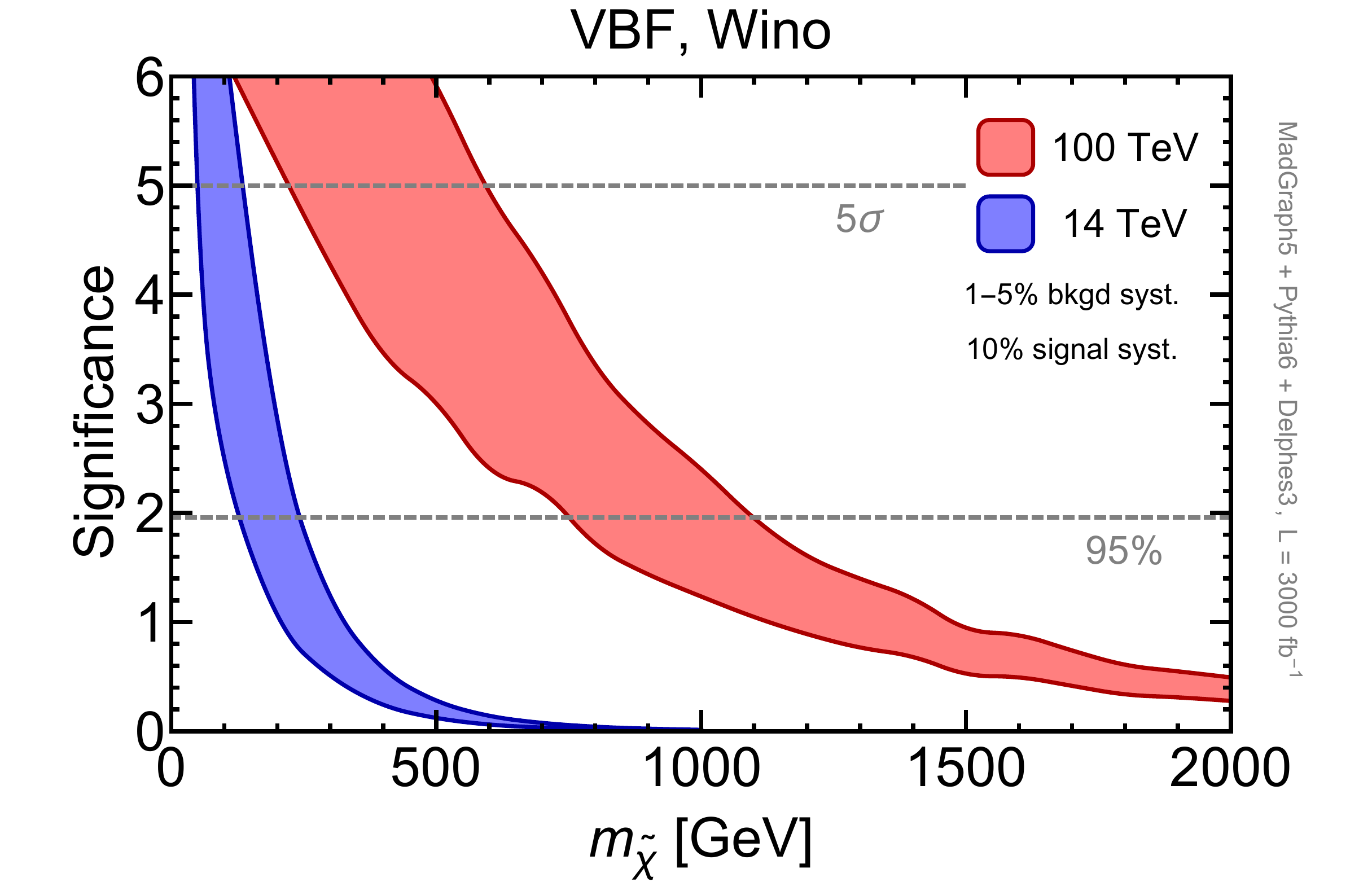}
  \includegraphics[width=0.49\textwidth]{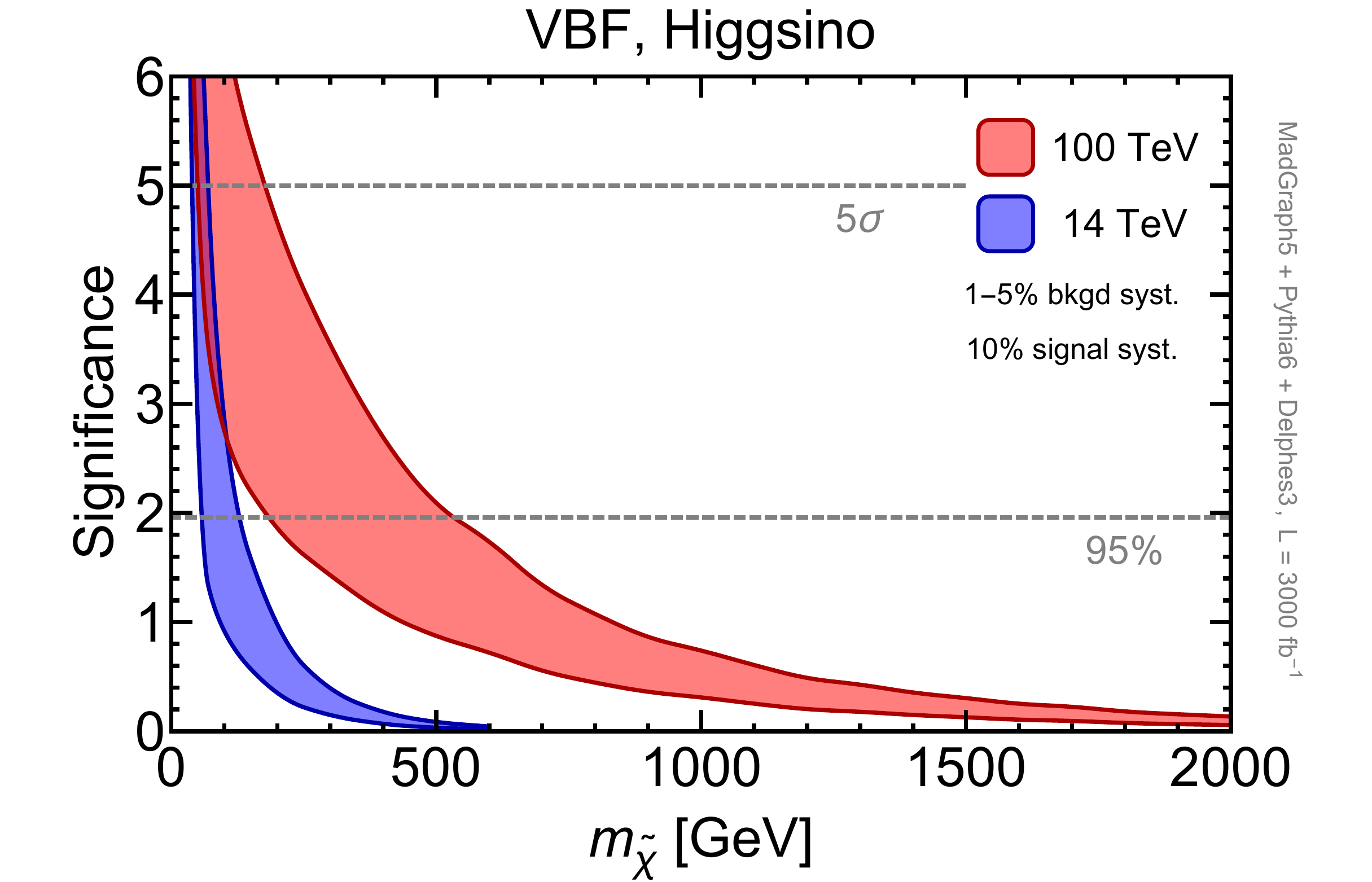}
  \caption{Wino/higgsino reach at 14 TeV (blue) and 100 TeV (red) on the left/right with 3000 fb$^{-1}$ of integrated luminosity.  The bands sweep out varying background systematics from $1-5\%$.}
  \label{fig:results}
\end{figure*}
%

%%%%%%%%%%%%%%%%%%%%%%%%%%%%%%%%%%%%%%%%%%%%%%%%%%%%%%%%%%%%%%%%%%%%%%%%%%%%%%%%%%%%%%%%%%%%%%%%%%%%%%%%%%%%%%%%%%%%%%%%
\section{Neutralino Results} \label{sec:results}

In this section, we discuss the results of our study. The mass reach for pure wino or higgsino at the 14 TeV LHC and a future 100 TeV collider with an integrated luminosity of $3000\text{ fb}^{-1}$ is presented in Figure~\ref{fig:results}. In all cases we calculate the significance using Eq.~(\ref{eq:significance}) and a fixed signal systematic uncertainty ($\gamma_S$) of $10\%$. Since it is usually the case that $S \ll B$, our results are not very sensitive to the particular choice of $\gamma_S$.  On the other hand, the background systematic uncertainty ($\gamma_B$) is varied from $1-5\%$, which gives rise to the bands in Figure~\ref{fig:results}. Additionally, we present the corresponding ranges in mass reach for exclusion ($95\%$) and discovery ($5 \sigma$) in Tables~\ref{table:14TeVresults} and \ref{table:100TeVresults}.

As can be seen by comparing Figure~\ref{fig:results} or Tables~\ref{table:14TeVresults} and \ref{table:100TeVresults} to previous studies of other search channels at 14 and 100 TeV ({\it e.g.} monojet or disappearing track)~\cite{Low:2014cba,Cirelli:2014dsa}, the vector boson fusion channel is generally weaker in its ability to discover or exclude winos/higgsinos with masses on the order of a TeV. However, VBF will still serve as a useful and necessary complementary probe in future searches to confirm the consistency of a potential wino/higgsino discovery in monojet or disappearing track processes.

As mentioned, for the 100 TeV studies, we extended the calorimeter coverage up to $|\eta|=7$, anticipating that future detectors may extend to $|\eta| \lesssim 6-7$.  In this sample, $\lesssim 2\%$ of jets have $|\eta| \geq 5.5$.  This is in contrast with the vector boson fusion signal from strong electroweak symmetry breaking where a significant fraction of jets can have $|\eta| \geq 5.5$~\cite{Ismail:future}.

The reason for the difference comes from the coupling of pure electroweakinos to gauge bosons.  In strong electroweak VBF the dominant process at high energies is the scattering of longitudinally polarized gauge bosons, {\it i.e.} two goldstone bosons scattering to two goldstone bosons.  Pure electroweakinos, on the other hand, do not couple directly to goldstone bosons\footnote{Electroweakinos can couple to goldstone bosons when they are mixed due to their higgsino-gaugino-higgs couplings.} so in VBF production of winos and higgsinos we can expect the interacting gauge bosons to be dominantly transversely polarized.

When longitudinally polarized gauge bosons are emitted off of quarks they have similar energy, but lower $p_T$ than transversely polarized modes.  This forces the forward jets to be pushed to higher pseudorapidity.  With electroweakinos in the final state, longitudinal modes are not present in the goldstone limit which means the jets have relatively high $p_T$ and lower pseudorapidity, $|\eta|\lesssim5$.  From this we conclude that extended calorimetry does not significantly impact this search.

\begin{figure}[tb]
  \includegraphics[width=0.49\textwidth]{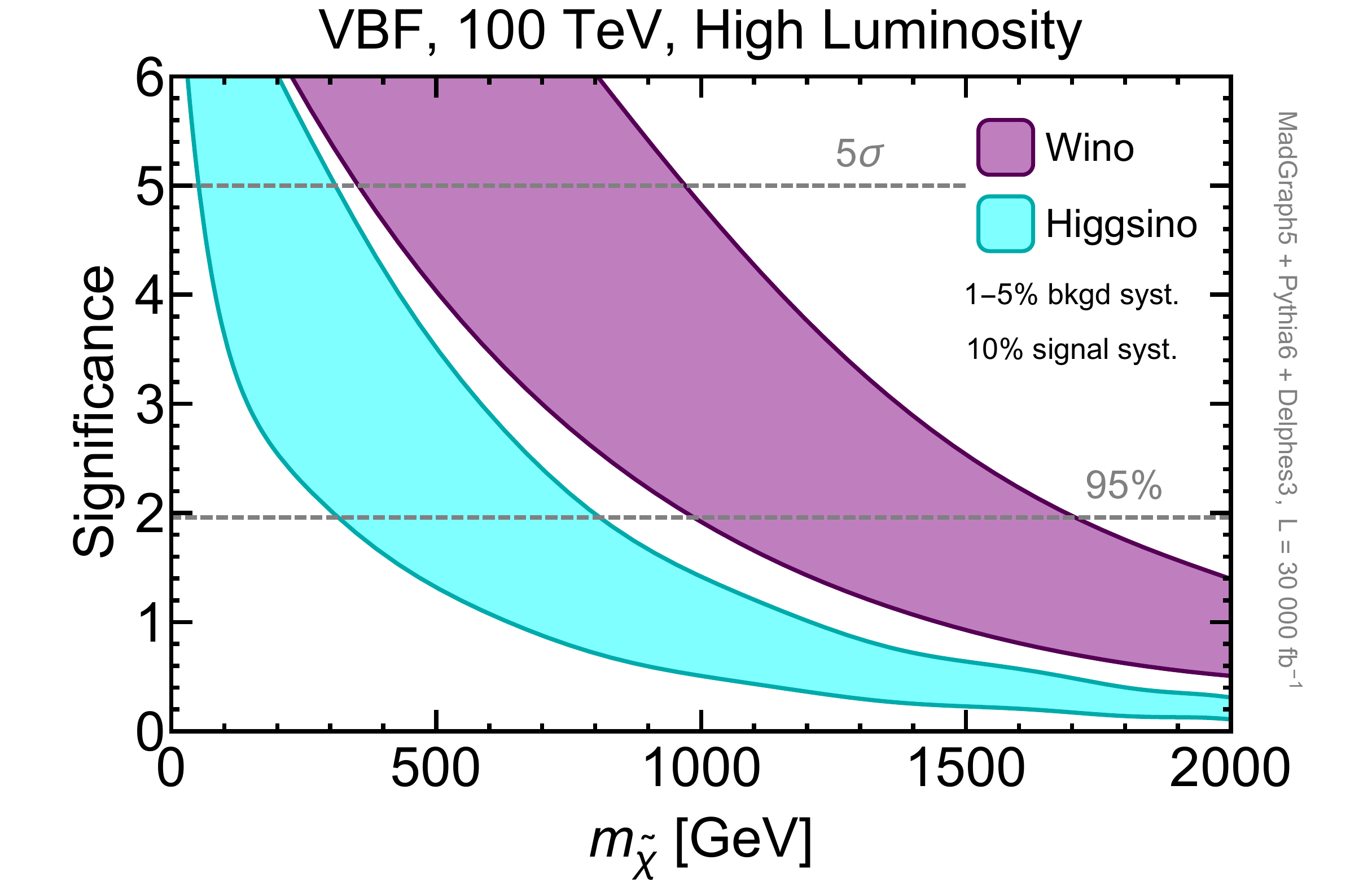}
  \caption{Wino/higgsino reach at 100 TeV with 30 ab$^{-1}$ of integrated luminosity.  The bands sweep out varying background systematics from $1-5\%$.}
  \label{fig:highlumi}
\end{figure}

Finally, to evaluate the impact of integrated luminosity on this search we perform an extrapolation of our results to a dataset of 30 ab$^{-1}$.  Rather than generating an increased dataset for these, we simply multiply our existing set accordingly.  Compared to the third and fourth columns of Table~\ref{table:cuts}, the cuts for $p_T(j_{\text{tag}})$, $p_T(j_{\text{veto}})$, and $\met$ are increased to 100 GeV, 100 GeV, and $1100 - 3000$ GeV, respectively. For both wino and higgsino, $M(j_1,j_2)$ and $\Delta \phi$ cuts of 10 TeV and 2 are used.  The other cuts are unchanged compared to the 3000 fb$^{-1}$ analysis. Since the missing energy cut is the most sensitive, we extrapolate the distribution smoothly to higher values to avoid effects from finite statistics.  The results are shown in Figure~\ref{fig:highlumi}.

%%%%%%%%%%%%%%%%%%%%%%%%%%%%%%%%%%%%%%%%%%%%%%%%%%%%%%%%%%%%%%%%%%%%%%%%%%%%%%%%%%%%%%%%%%%%%%%%%%%%%%%%%%%%%%%%%%%%%%%%
\section{General Results} \label{sec:general}

Our analysis is aimed at models where dark matter is minimally introduced as the electrically neutral component of an electroweak $n$-tuplet ($n=2,3$ corresponding to higgsino and wino, respectively), which can then be produced in VBF processes via its interactions with gauge bosons.  The signal of VBF and missing energy is therefore quite generic, and futhermore appears in many models of new physics.  For instance, if there is a new gauge singlet that is stable on collider time scales and possesses significant interactions with the Higgs, then VBF is an effective way to probe this type of model (see {\it e.g.}~\cite{Craig:2014lda}).

\begin{figure*}[t]
  \includegraphics[width=0.49\textwidth]{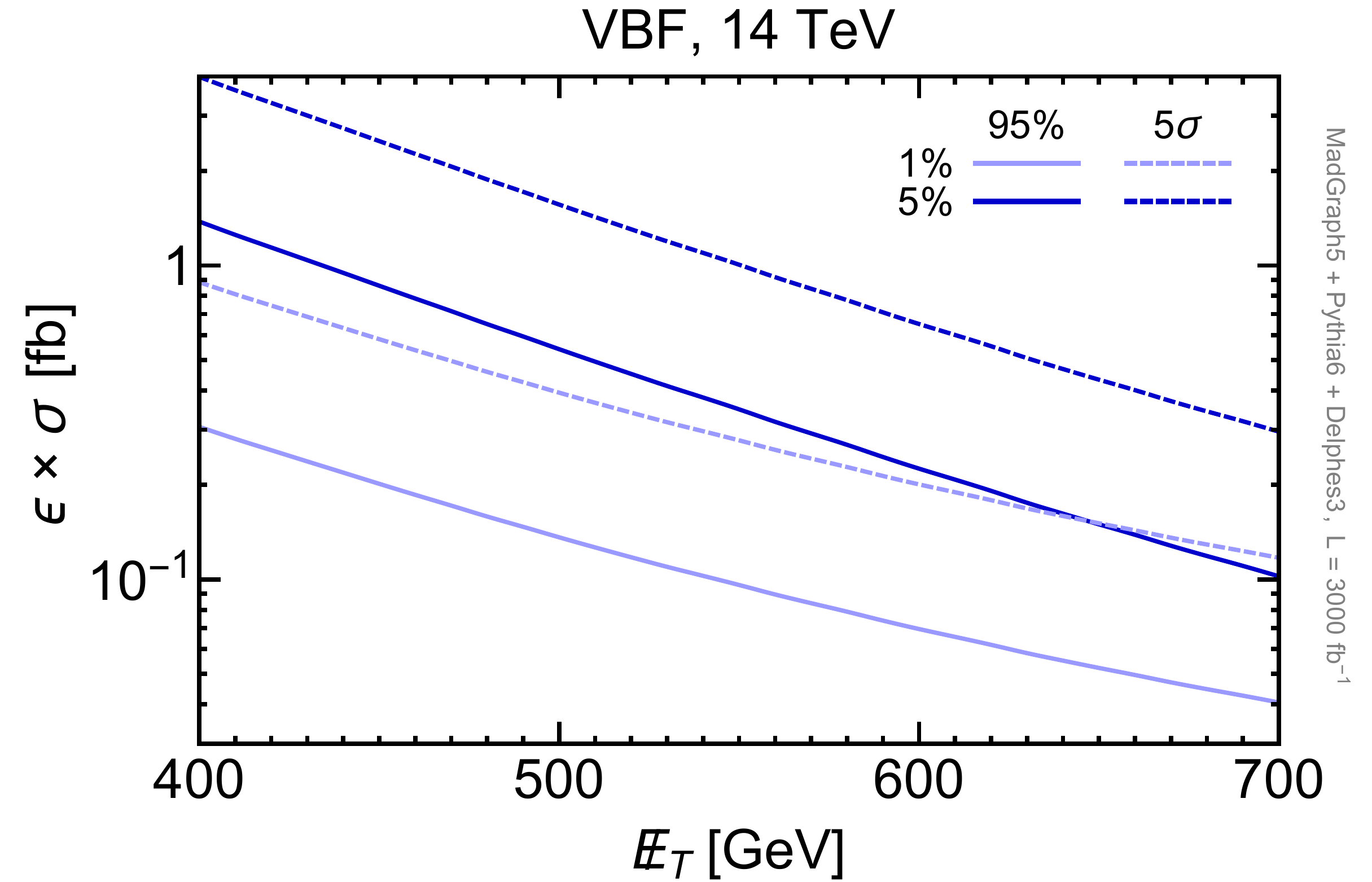}
  \includegraphics[width=0.49\textwidth]{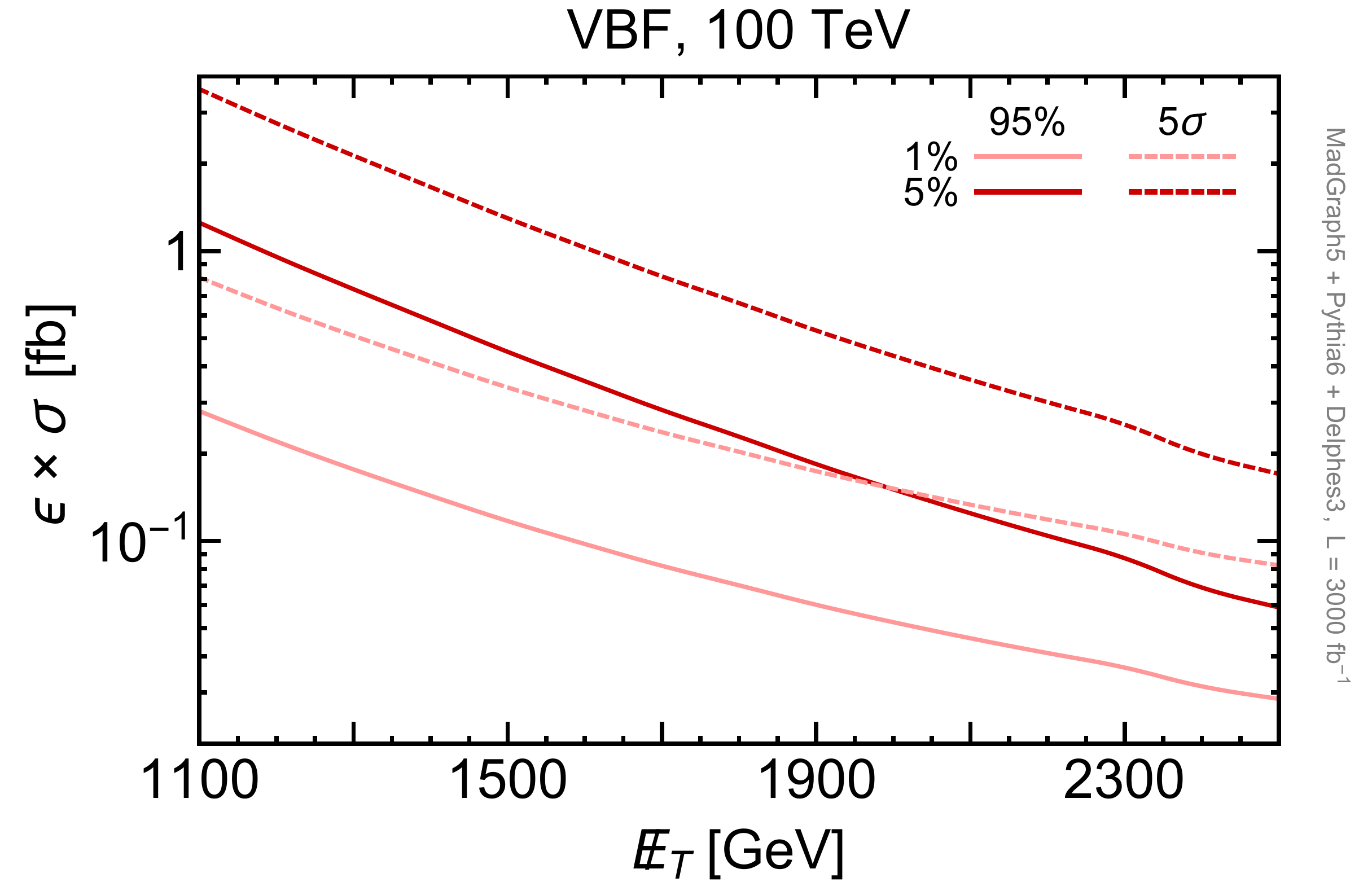}
  \caption{Rates that can be excluded ($95 \%$) (solid) or discovered ($5 \sigma$) (dashed) at 14 TeV (left) and 100 TeV (right) for background systematics of 1 or 5 \% as a function of the missing energy cut. All other cuts have been included, and in particular are chosen to be the midpoint between the optimized cuts for wino and higgsino. A signal systematic uncertainty of 10\% and an integrated luminosity of 3000 fb$^{-1}$ are assumed.}
  \label{fig:genresults}
\end{figure*}

In light of the broad applicability of the VBF with missing energy channel, and the care that must be taken in generating the backgrounds, we provide exclusion and discovery contours for cross-sections vs. missing energy cuts.  Figure~\ref{fig:genresults} displays the cut efficiency times cross-section after all cuts from Table~\ref{table:cuts} have been applied (and fixed to the average values between wino and higgsino) as a function of the cut on missing energy.  Given a simulated signal, these plots can be utilized in a simple fashion to obtain sensitivity for that model.  To do so, one applies Table~\ref{table:cuts}'s cuts to the signal and compares to $\epsilon \times \sigma$ to exclude/discover the signal at 14 or 100 TeV.

%%%%%%%%%%%%%%%%%%%%%%%%%%%%%%%%%%%%%%%%%%%%%%%%%%%%%%%%%%%%%%%%%%%%%%%%%%%%%%%%%%%%%%%%%%%%%%%%%%%%%%%%%%%%%%%%%%%%%%%%
\section{Conclusions} \label{sec:conclusions}

Producing and studying dark matter remains one of the main goals of the LHC and constitutes one of the primary targets for a future 100 TeV $pp$ collider. Meanwhile, the vector boson fusion channel is an important component of new physics searches, and here we analyze its relevance to dark matter. In studying various dark matter scenarios, supersymmetry provides a very useful set of examples.  We presented a thorough study in the cases where the higgsino (electroweak doublet) and the wino (electroweak triplet) constitute the dark matter.

In this work we analyzed the reach in the VBF plus missing energy channel.  We found that the reach is 240 GeV for winos and 125 GeV for higgsinos at 14 TeV.  Going to 100 TeV, the respective sensitivity increases to 1.1 TeV and 530 GeV.  While VBF is not the discovery channel for electroweak dark matter, if hints of dark matter were observed in a monojet search, the VBF channel would provide a crucial verification.  This is analogous to the Higgs discovery in which all available channels need to be looked at to fully understand its properties.

Since missing energy is a generic signature of models of new physics with dark matter candidates, in Section~\ref{sec:general} we used the simulated backgrounds to set model-independent limits on cross-sections in vector boson fusion.

Finally, we investigated the impact of extended calorimetry on the neutralino reach and found that it does not impact this search. Compiling a list of requirements for proposed detectors of a 100 TeV collider is important: as evidenced in this study, the search for wino or higgsino dark matter can only be touched upon at the LHC and would benefit immensely from an  increase to 100 TeV.

%%%%%%%%%%%%%%%%%%%%%%%%%%%%%%%%%%%%%%%%%%%%%%%%%%%%%%%%%%%%%%%%%%%%%%%%%%%%%%%%%%%%%%%%%%%%%%%%%%%%%%%%%%%%%%%%%%%%%%%%
\begin{acknowledgments}
The authors would like to thank Nathaniel Craig for useful discussions and Xerxes Tata for helpful comments.

This work was supported in part by the Kavli Institute for Cosmological Physics at the University of Chicago through grant NSF PHY-1125897 and an endowment from the Kavli Foundation and its founder Fred Kavli.  ML is partially supported by NSERC of Canada and LTW is supported by DOE grant DE-SC0003930.  Computations for this paper were performed on the Midway and Orion clusters supported by PSD Computing at the University of Chicago.
\end{acknowledgments}

\bibliography{higgsino}
\end{document}